# Unusual magneto-transport from Si-square nets in topological semimetal HfSiS

Nitesh Kumar, [1] Kaustuv Manna, [1] Yanpeng Qi, [1] Shu-Chun Wu, [1] Lei Wang, [1,2] Binghai Yan, [1] Claudia Felser, [1] and Chandra Shekhar [1,*]

[1]*Max Planck Institute for Chemical Physics of Solids, 01187 Dresden, Germany.*
[2]*Department of Power and Electrical Engineering, Northwest A&F University, Yangling, Shaanxi 712100, China.*

**Abstract:** The class of topological semimetals comprises a large pool of compounds. Together they provide a wide platform to realize exotic quasiparticles for example Dirac, nodal line Dirac and Weyl fermions. In this letter, we report the Berry phase, Fermi surface topology and anisotropic magnetoresistance of HfSiS which has recently been predicted to be a nodal line semimetal. This compound contains large carrier density, higher than most of the known semimetals. Massive amplitudes of de Haas-van Alphen and Shubnikov-de Haas oscillations up to 20 K in 7 T assist us in witnessing nontrivial π-Berry phase which is a consequence of topological Dirac–type dispersion of bands originating from the hybridization of $p_x + p_y$ and $d_{x2-y2}$ orbitals of square–net plane of Si and Hf atoms, respectively. Furthermore, we establish the 3D Fermi-surface which consists of very asymmetric water caltrop-like electron and barley seed-like hole pockets which account for the anisotropic magnetoresistance in HfSiS.

**Introduction:** In graphene, carbon atoms are arranged in honeycomb lattice that facilitates a linear relation between energy and momentum, Dirac-type dispersion [1] near the Fermi energy. Additionally, a negligible spin orbit coupling (SOC) in carbon atom maintains the gapless Dirac crossing [2]. It has recently been realized that many other families of compounds also have such Dirac-type crossing in bulk and they, however, possess the square-net arrangement of atoms namely Bi square-net in $A$MnBi$_2$, where $A$ = Sr, Ba, Ca, and Eu [3-5], $R_2$O$_2$Bi [6,7] and $R$AgBi$_2$ [8,9], where $R$ = rare earth or Y, Se and Te square-net in selenene, tellurene and $R$Te$_3$ [10,11], arsenic square-net in $R$CuAs$_2$ [12] *etc*. Evolution of such a Dirac–type bands crossing upon the inclusion of SOC normally leads to a negligible small gap in graphene [2] while it is large in compounds containing heavy atoms like Bi. Furthermore, depending on the crystal structure and symmetries present in the compound some Dirac crossings are protected against the opening of the gap [13,14]. In nodal line semimetals, the valence and conduction bands, however, cross along in a particular direction in momentum space forming a drumhead like surface states [15-18]. Such compounds show extraordinary transport properties e.g. low effective mass of carrier, extremely high magnetoresistance (MR), extremely high mobility *etc*. [13,14,19-23]. Very recently, a group of ternary materials with general formula $WHM$ ($W$= Zr, Hf, or La; $H$ = Si, Ge, Sn, or Sb; $M$ = O, S, Se and Te) have been predicted as 2D topological insulators [24]. All these compounds have PbFCl-type crystal structure (space group $P4/nmm$) as shown in the Fig. S1(a) for HfSiS wherein Si atoms (in general $H$ atoms) arrange in square–net manner. This series of compounds exhibit two interesting features. First, a nontrivial nodal line states are protected by glide-mirror symmetry below the Fermi energy (-0.5 eV in HfSiS) [15-18]. Second, several other trivial Dirac-type crossings at Fermi level originate from $p_x + p_y$ orbitals of square-net Si atoms and $d_{x2-y2}$ orbitals of Hf atoms in HfSiS. These Dirac crossings are not protected by symmetries, for example a gap opening of 30 meV and 20 meV is observed in ZrSiO [24] and ZrSiS [25], respectively because of SOC.

The square-net arrangement of Si atoms in the *ab*-plane imparts layered structure in these compounds which results in the plate-like crystals growth [Fig. S1(b)]. Among the predicted *WHM* series, only some Zr compounds have been studied ZrSi(S,Se,Te) [19-23,25-27] and all the compounds show very similar transport properties. Surprisingly, the noteworthy experimental application of quantum oscillations to construct 3D Fermi surface (FS) is still missing. It is more interesting to investigate HfSiS, a sister compound of ZrSiS, which has a larger spin-orbit coupling and hence shows a clear unexpected Dirac-node arc [28]. In this study, we report Berry phase, MR and the Fermiology of HfSiS. We find that HfSiS shows $\pi$–Berry phase, low effective mass of carriers, high MR and mobility that reveal the Dirac–like dispersion of bands which indeed control the transport properties. Highly anisotropic MR is due to the asymmetric electron and hole pockets and large mass anisotropy. We further believe that the transport characteristics in HfSiS and related compounds significantly depend on the Dirac-like dispersion originated from the hybridized $p_x + p_y$ orbitals of square-net Si atoms and $d_{x2-y2}$ orbitals of Hf atoms near the Fermi level rather than on the nodal line states which lie deeper in the valence band [29].

**Results and discussion:** Thin plate–like crystals of HfSiS were grown by chemical vapor transport method. The details about the synthesis route and characterization were described in the supplementary information (SI). From Fig. S1(d), the resistivity, $\rho_{xx}$ (T) of HfSiS without magnetic field initially increases with $T^2$ up to 100 K after that it shows linear dependency with T like in a metal. The $\rho_{xx}$ changes an order of magnitude from 2 K (3.1 $\mu\Omega$ cm) to 300 K (33 $\mu\Omega$ cm) and we found similar values in different batches of crystals. A dominating p-type charge carrier density of $6\times10^{20}$ cm$^{-3}$ at 2 K was obtained from Hall resistivity which does not show a strong temperature dependence like in other low carrier semimetals [30].

To derive the physical parameters of the charge carriers from quantum oscillations, we measured isothermal magnetization of HfSiS up to 7T in $B \parallel [100]$ and $B \parallel [001]$ orientations at different temperatures as shown in Figs. 1(a) and (d), respectively. Magnetization exhibits a clear de Haas-van Alphen (dHvA) quantum oscillations starting from 1 T indicating cleanness of the compound and low effective masses of charge carriers. From an instantaneous view in $B \parallel [100]$, these oscillations seem to involve mainly two frequencies; where the one with larger frequency and small amplitude is superimposed on the second one with small frequency and huge amplitude. However, the total number of frequencies in an oscillation is seen in its fast Fourier transformation (FFT). After subtracting a smooth background [see Fig. S2(b)] from the measured data at each temperature, the periodic oscillations are visible in $1/B$ up to 20 K [Fig. 1(b)]. Oscillations at the intermediate temperature range can be seen in Fig. S2. The corresponding FFT shows four frequencies [inset of Fig. 1(c)] namely the hole pocket $\alpha = 13.5$ T and electron pockets $\beta = 120$ T, $\gamma = 124$ T and $\eta = 138.5$ T which are compatible with the band structure [24]. These frequencies are directly linked to the extremal area of the pockets ($A_F$) by the Onsager relation, $F = (\Phi_0/2\pi^2) A_F$, where $\Phi_0 = 2.068\times10^{-15}$ Wb. $A_F$ corresponding to the pockets $\alpha$ and $\eta$ are 0.0013 Å$^{-2}$ and 0.013 Å$^{-2}$, respectively which is consistent with the existence of tiny pockets near the Fermi energy. Many important characteristics of actively involved charge carriers in quantum oscillations can further be determined from the quantitative analyses of observed dHvA oscillations. In general, the oscillatory behavior of magnetization is describes by Lifshitz-Kosevich (LK) formula [31] accounting the Berry phase in a 3D Dirac system:

$$\Delta M \propto -B^{\frac{1}{2}} R_T R_D \sin\left[2\pi\left(\frac{F}{B} + \gamma - \delta\right)\right] \qquad (1)$$

Where $R_T = 14.69 m^* T/B \sinh(14.69 m^* T/B)$, $R_D = \exp(-14.69 m^* T_D/B)$, $m^*$ is effective mass, $T_D$ is Dingle temperature, $(2\pi^2 k_B m_0)/(\hbar e) = 14.69$, and $m_0$ is the bare mass of electron. $R_T$ and $R_D$ are related to the broadening in Landau levels (LL) due to temperature effect in Fermi-Dirac distribution and electron scattering, respectively. The oscillatory behavior of $\Delta M$ is characterized by the sine term along with an additional phase factor $(\gamma - \delta)$, in which $\gamma = \frac{1}{2} - \phi_B/2\pi$ and $\phi_B$ is the Berry phase. However, an oscillatory behavior in resistivity is characterized by the cosine term. The phase shift $\delta$ is determined by the dimensionality of the Femi surface and it is 0 and $\pm 1/8$ ("+" for hole pocket and "−" for electron pocket) for the 2D and 3D cases, respectively. From the above LK formula, the value of $m^*$ can be determined through the fit of the temperature dependence of the oscillation amplitude with the thermal damping factor $R_T$, as shown in Fig. 1(c). We find the values of $m^*_\alpha = 0.08$ and $m^*_\eta = 0.17$, which are very low and are likely to originate from the Dirac–type dispersion of bands. Further, the Dingle temperature $T_D$ is determined to be ~ 3.5 K and ~ 6.5 K for $\alpha$ and $\eta$ pockets, respectively and corresponding quantum relaxation times, $\tau_q$ (= $\hbar/(2\pi k_B T_D) = 1.22 \times 10^{-12}/T_D$) are $3.5 \times 10^{-13}$ s and $1.9 \times 10^{-13}$ s. The quantum mobility, $\mu_q$ (= $e\tau_q/m^*$) is estimated to be $7.7 \times 10^3$ cm$^2$/Vs for $\alpha$ pocket and $2.0 \times 10^3$ for $\gamma$ pocket.

We also measured the magnetization isotherms up to 20 K for $B \parallel [001]$ and the data of the whole temperature range can be found in Fig. S3. In this geometry also, HfSiS shows strong dHvA oscillations [Fig. 1(d)] but their amplitudes are smaller as compare to $B \parallel [100]$ [Fig. 1(a)] due to curvature factor. This gives a hint about the elongated Fermi pockets in [001]; the detailed Fermiology will be discussed in the following sections. After a detailed analysis similar to $B \parallel [100]$, these oscillations are found to comprise of two frequencies namely $\varepsilon$ (31 T) and $\xi$ (264 T) [inset of Fig. 1(c)]. The corresponding extremal area of pockets $\varepsilon$ and $\xi$ are 0.003 Å$^{-2}$ and 0.025 Å$^{-2}$, respectively. In contrast to $B \parallel [100]$, both pockets are of hole type and exhibit same effective mass of 0.13. More detailed information concerning the pockets are given in the table-1. The low value of $m^*$ and high values of $\tau_q$ and $\mu_q$, all together indicate the evidence of Dirac-like fermions from multiple linear bands. Furthermore, under the certain conditions such as the linear band-contact line in the momentum space, the charge carriers acquire an extra geometrical phase in the oscillations which is known as the Berry phase [32]. Now, we focus our results on the Berry phase which gives the in-depth view of Dirac-like dispersion of individual bands. Usually, it is a challenging procedure to index Landau levels when multiple frequencies are involved in oscillations. The Landau fan diagram is plotted between $n$ and $1/B$ from Lifshitz-Onsager relation

$$n = \frac{F}{B} + \gamma - \delta \pm \frac{1}{4} \qquad (2)$$

Where $F$ is a fundamental frequency and an additional phase of +¼ (−¼) corresponds to the minima (maxima) of dHvA oscillation peaks. However, besides constructing Landau fan diagram, we used LK equation to calculate the Berry phase. For the $\alpha$ band ($F = 31$ T, $m_\alpha = 0.08$, $T_D = 3.5$ K) taken as an example; the sine function with the Dingle factor of equation (1) is fitted to the oscillations spectra as shown in Fig. 2(a) and best fitting gives the value of $|\gamma - \delta|$ equal to 0.93. By considering 3D hole pocket ($\delta = 1/8$), this directly gives the value of $\phi_B$ as 1.1π which is close to π-Berry phase indicating a non-trivial band. On the other hand, the Berry phase information related to $\beta$, $\gamma$ and $\eta$ bands are not possible to determine accurately at this moment because their frequencies are very close to each other which are also reflected by beating patterns in oscillations [Fig. 2(b)] and their related amplitudes sharply decrease with temperature in FFT [Fig. 1(c)]. These behaviors restrict us to find the effective mass and Dingle factor and hence the Berry phase related to the pockets $\eta$, $\beta$ and $\gamma$. Now for the case $B \parallel [001]$, HfSiS shows very clean dHvA oscillations unlike to $B \parallel [100]$ and no beating patterns are observed. Therefore, LK fit [Fig. 2(d)] gives the values of $|\gamma - \delta|$ equal to 0.94 and 0.89 for $\varepsilon$ and $\xi$, respectively revealing that the phase is $1.1\pi$ for $\varepsilon$ ($\delta = 1/8$ for hole pocket) and $0.47\pi$ for $\xi$ ($\delta = -1/8$ for electron pocket). These values indicate the non-trivial and trivial Berry phase for $\varepsilon$ and $\xi$, respectively. It is interesting to note that HfSiS behaves electronically very similar to SrMnBi$_2$. SrMnBi$_2$ also exhibits non-trivial $\pi$ Berry phase from the Dirac cones (gapped marginally by SOC) near $E_F$. Although both the compounds comprise of square net arrangement of Si/Bi, in case of SrMnBi$_2$ [33], states near $E_F$ are dominated by $p$-orbitals of square net of Bi in contrast to HfSiS where the states near $E_F$ originate from the hybridization of Si square net $p$ orbitals and d orbitals of Hf. By comparing compounds of HfSiS series with structurally similar LaSbTe (no $p$-$d$ hybridization) it has been seen theoretically that such hybridization of $p$ and $d$ orbitals barely changes the electronic properties [24]. Furthermore, the nodal line states in HfSiS which is protected by non-symmorphic crystal symmetry hardly affect the electronic properties because they are situated far below $E_F$.

In the above, we discussed features of the electronic structure considering the maximal area of the Fermi pockets in the (110) and (011) planes. Due to the tetragonal symmetry along $c$-axis, (011) and (101) planes are equivalent in HfSiS. In order to construct the 3D Fermi surface experimentally we monitor maximal areas along different directions in addition to the crystallographic axes by measuring magneto-resistivity in rotating field. SdH oscillation frequencies were extracted from measured data as shown in Fig. S4. The rotation geometry of $\theta$ and $\phi$ define are as $\theta = 0°$ ($B \parallel$ [001]) to $\theta = 90°$ ($B \parallel [100]$) and $\phi = 0°$ ($B \parallel [001]$) to $\phi = 90°$ ($B \parallel [110]$) as shown in the inset of the Fig. 3(a). After carefully following the peak positions in Figs. 3(b), (c), we find some interesting facts. The peak, $\xi$ varies almost in the similar way in both the directions and surprisingly it disappears at $\theta > 60°$ while the frequency of $\varepsilon$ slightly increases with the angle. The last tracked frequency of $\xi$ is 345 T at $\phi$, $\theta = 50°$. At smaller values of $\phi$ and $\theta$, $\xi$ follows 1/cosine behavior, however, it shows deviation after 40° indicating quasi 2D nature of the pocket similar to Zr compounds [19-22,25,27]. Furthermore, the frequency of $\alpha$ eventually appears at 22 T for $\theta = 30°$ and 35 T for $\phi = 30°$ while showing a decreasing trend with increasing $\theta$ and $\phi$. The large shifting, appearance and disappearance of peaks with rotating magnetic fields describe the asymmetric feature of the Femi pockets in HfSiS. In the entire *WHM* family, the orbitals $d_{x2-y2}$ of *W* and $p_x+p_y$ of *H* mainly contribute at Fermi level and effect of p–orbitals of *M* is negligible. We have traced $\theta$ and $\phi$ dependence of frequencies with full angular rotation as described above and plotted in Fig. 4(a) in open circles along with theoretically calculated values (blue crosses for electrons and red crosses for holes). This clearly shows the frequency variation in different branches due to the anisotropic Fermi pockets. From *ab-initio* calculations, we constructed the bulk 3D Fermi pockets as shown in Fig 4(b). A spherical and four water caltrop–like shape electron pockets (blue) are located at $k_z = 0$ and $k_z = 0.5$, respectively while four barley-like shape hole pockets (red) at $k_z = 0.5$ in the first Brillion zone. Figs. 4(c) and 4(d) show the polar plots of $\rho_{xx}$ at 2 K and 50 K in different fields, respectively in a configuration where *I* is always $\perp B$ in order to maintain a uniform Lorentz force contribution. We observe a large change in the resistivity as a function of the field direction. It is a direct consequence of asymmetric Fermi surfaces involved in HfSiS. The ripples observed in 9 T data at 2 K is a result of strong quantum oscillations.

In conclusion, we determined the 3D Fermi-surface topology of the HfSiS and it consists of asymmetric electron and hole pockets. A very low carrier effective mass and $\pi$–Berry phase are originated from Dirac–type dispersion of bands. Such linear dispersion near the Fermi level is the direct consequence of the interplay between $p_x + p_y-$ and $d_{x2-y2}$–orbitals from Si square nets and Hf atoms respectively. The Fermi surface of HfSiS is quite anisotropic which directly reflects in the large anisotropy in magnetoresistance. Our results provide an insight into the compounds containing square−net atomic arrangements with hybridized *p-d* orbitals and how they are correlated electronically to square net compounds without such hybridization.

**Acknowledgements**: We acknowledge financial supports from the Max Planck Society and ERC Advanced Grant (291472 Idea Heusler.

**Figure captions:**

FIG. 1. For B || [100], (a) Isothermal magnetization at different temperatures, (b) amplitude of dHvA after subtraction of background, (c) temperature dependent FFT amplitudes. Inset of (c) shows FFT of oscillations mentioned in (b) that depicts four frequencies $\alpha$, $\beta$, $\gamma$ and $\eta$ at 13.5 T, 120 T, 124 T and 138.5 T, respectively. Corresponding data for B || [001] case in (d), (e), and (f). In this case, FFT shows only two frequencies $\varepsilon$ and $\zeta$ at 31 T and 264 T, respectively.

FIG. 2. Fitting of LK formula to determine Berry phase for (a) $\alpha$ when B || [100], (b) $\varepsilon$ and (c) $\xi$ when B || [001], where black line is measured data and red line is fit.

FIG. 3. (a) Magneto-resistance in different angles and their corresponding FFTs in (b) for $\theta = 0^\circ$ to $90^\circ$. (c) FFTs of SdH oscillations for $\phi = 0^\circ$ to $90^\circ$. Rotation directions $\theta$ and $\phi$ are defined with respect to B in the inset of (a).

FIG. 4. (a) Experiential frequencies (open circle) and theoretically calculated frequencies from SdH oscillations (blue cross for electron pocket and red cross for hole pocket), (b) 3D Fermi pockets (blue for electron and red for hole) in the first Brillion zone, (c) and (d) polar plots of $\rho_{xx}$ at 2 K and 50 K in various fields when $I$ always $\perp B$.

TABLE 1. The derived parameter from dHvA oscillations for HfSiS. $F$, oscillation frequency; $m^*$, effective mass; $T_D$, Dingle temperature; $\tau_q$, quantum relaxation time; $\mu_q$, quantum mobility; $\phi_B$, Berry phase.

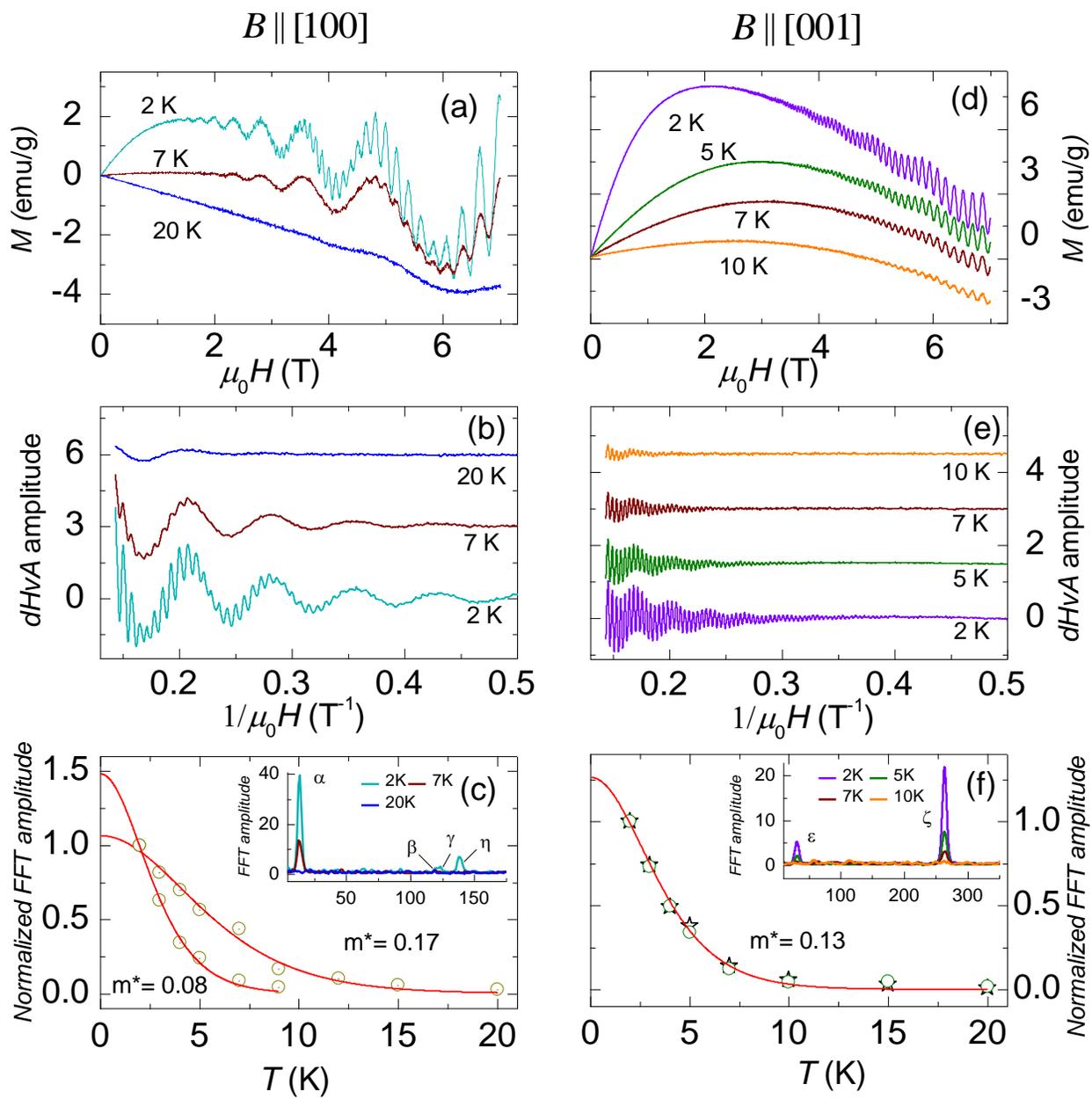

FIG. 1

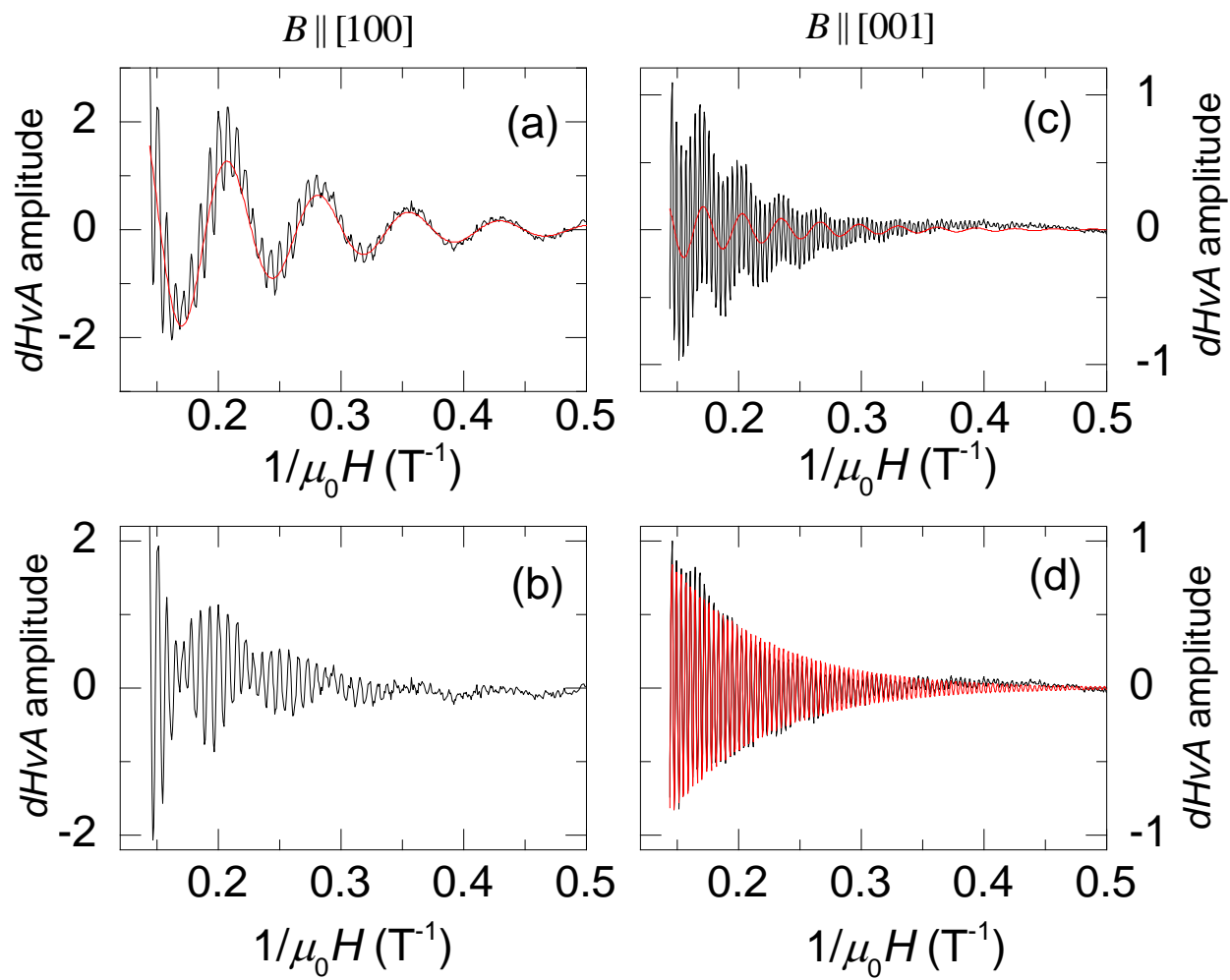

FIG. 2

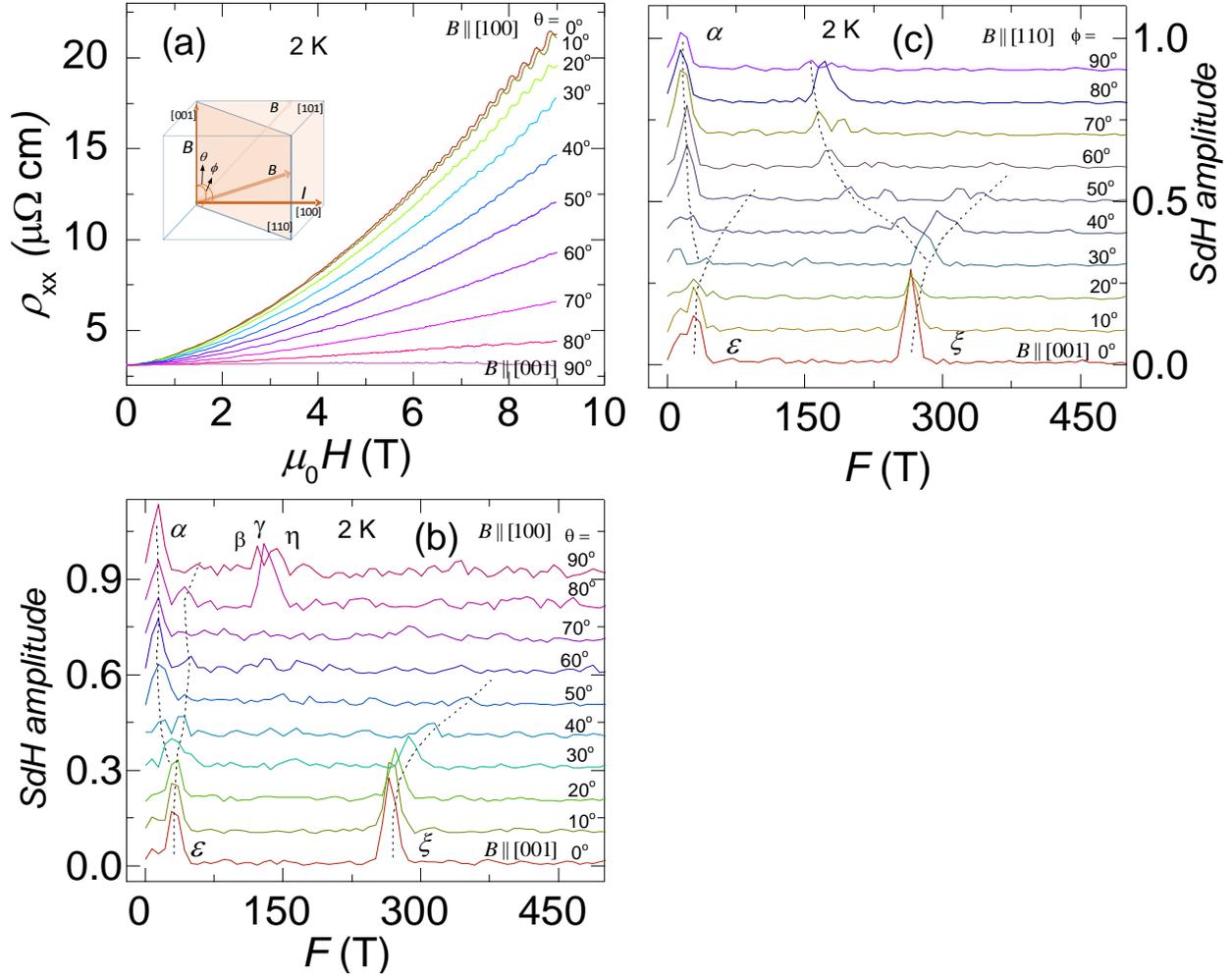

FIG. 3

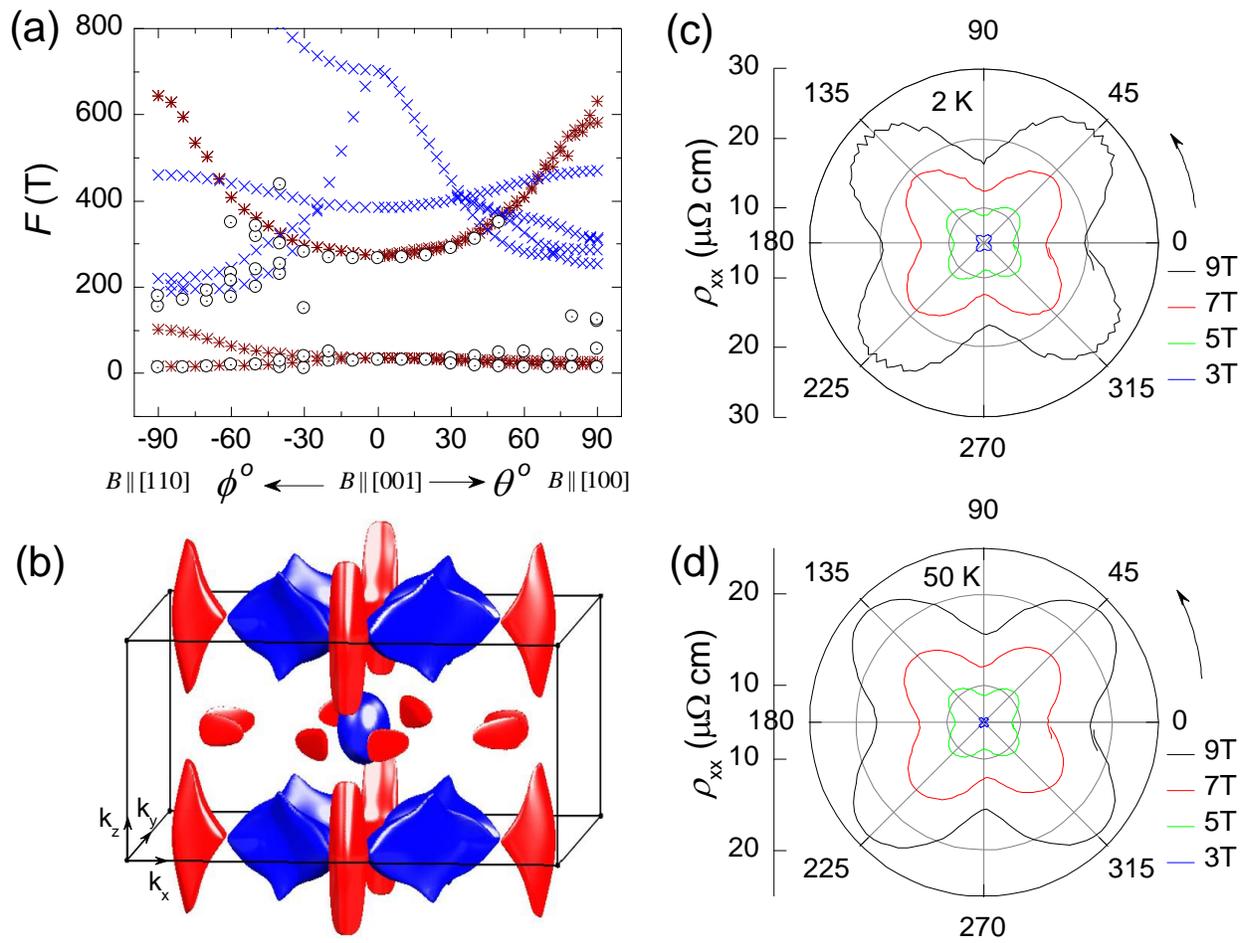

FIG. 4

TABLE 1.

|  | $F$ (T) | $m^*/m_0$ | $T_D$ (K) | $\tau_q$ (s) | $\mu_q$ (cm$^2$V$^{-1}$s$^{-1}$) | $\phi_B$ |
|---|---|---|---|---|---|---|
| $B \parallel [100]$ | 13.5 | 0.08 | 3.5 | $3.5 \times 10^{-13}$ | $7.7 \times 10^3$ | $1.1\pi$ |
|  | 120 |  |  |  |  |  |
|  | 124 |  |  |  |  |  |
|  | 138.5 | 0.17 | 6.5 | $1.9 \times 10^{-13}$ | $2.0 \times 10^3$ |  |
| $B \parallel [001]$ | 31 | 0.13 | 5.9 | $2.1 \times 10^{-13}$ | $2.8 \times 10^3$ | $1.1\pi$ |
|  | 264 | 0.13 | 6.2 | $2.0 \times 10^{-13}$ | $2.7 \times 10^3$ | $0.47\pi$ |

# Supplementary Information

Unusual magneto-transport from Si-square nets in topological semimetal HfSiS

Nitesh Kumar, [1] Kaustuv Manna, [1] Yanpeng Qi, [1] Shu-Chun Wu, [1] Lei Wang, [1,2] Binghai Yan, [1] Claudia Felser, [1] and Chandra Shekhar [1,*]

[1]Max Planck Institute for Chemical Physics of Solids, 01187 Dresden, Germany.
[2]Department of Power and Electrical Engineering, Northwest A&F University, Yangling, Shaanxi 712100, China.

**Crystal growth, material characterizations and theoretical calculations:** Single crystals of HfSiS were grown by chemical vapor transport method using iodine ($I_2$) as a transport agent. First, the polycrystalline HfSiS powder was synthesized with high-purity elemental Hf piece, Si piece, S powder in a process described elsewhere (*1*). Second, the polycrystalline powder together with iodine were sealed in a quartz tube under vacuum. The sealed quartz tube was then kept in a gradient furnace for 10 days at 1100◦C, while the colder zone at 1000◦C. Shinny rectangular plate like crystals were obtained at the colder zone. Crystallographic directions were measured by Laue X-rays method. Magnetization and transport measurement were performed in Quantum Design Inc. SQUID-VSM and PPMS, respectively. Density-functional theory (DFT) implemented in the Vienna ab-initio simulation package (VASP) was adopted (*2, 3*). The generalized gradient approximation (GGA) with spin orbital coupling was included for the bulk Fermi surface calculations.

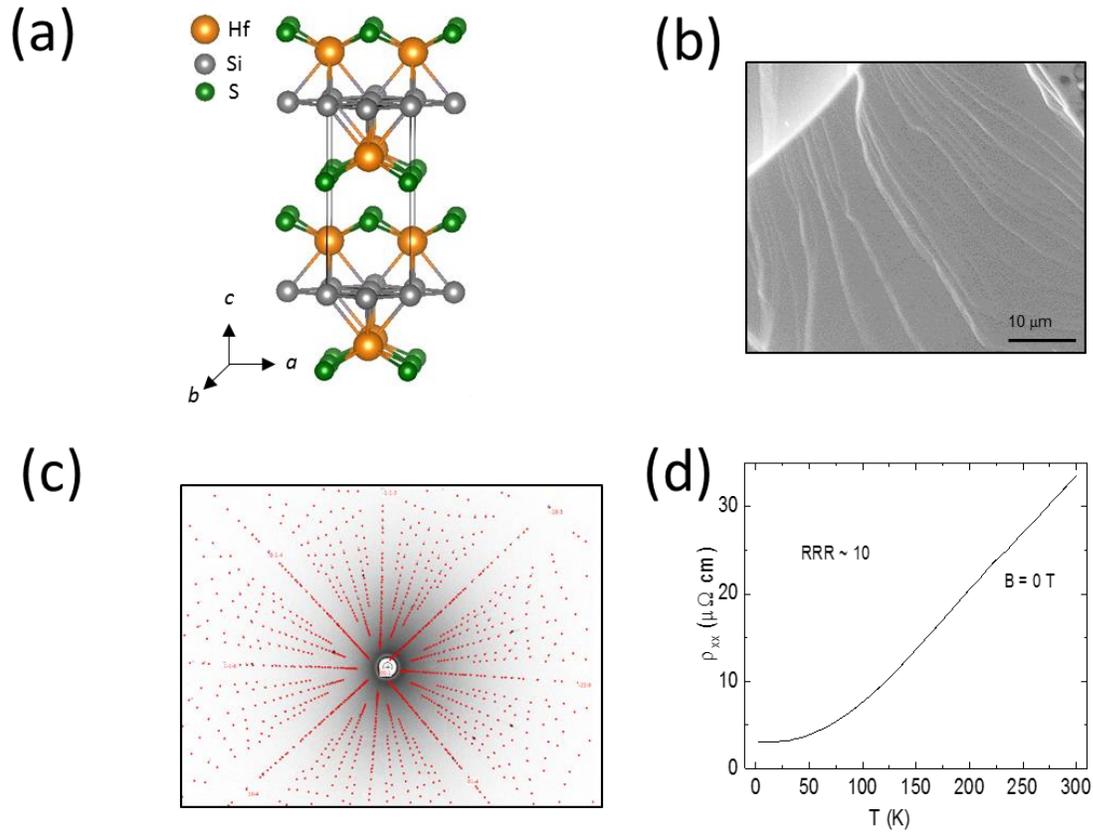

**FIG. S1** (a) Unit cell of HfSiS showing Si nets in gray color atomic lattice. (b) Scanning electron microstructural image depicting the formation of layered structures in (110) plane which is equivalent to Si nets plane in unite cell. (c) Single crystal Laue diffraction patterns in which x-rays beam is focused parallel to [001]. (d) Temperature dependent resistivity, $\rho_{xx}$ without magnetic field where current is || [100].

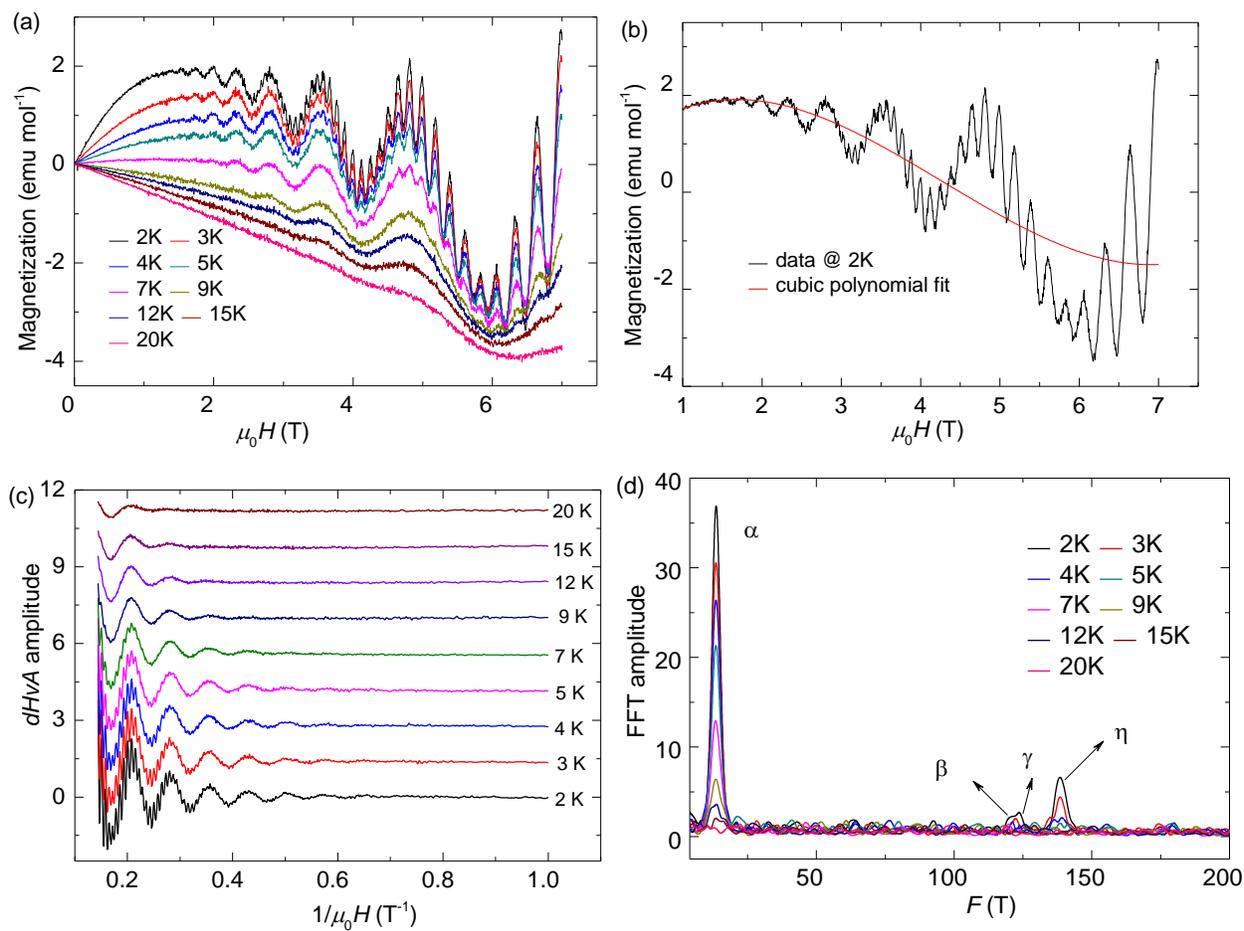

**FIG. S2.** B ∥ [100] and temperature range 2 – 20 K for (a) Isotherm magnetization. (b) Measured magnetization data at 2 K with cubic polynomial fit. (c) AdHv oscillations after background subtracted from (b). (d) Fast furrier transforms of data (c).

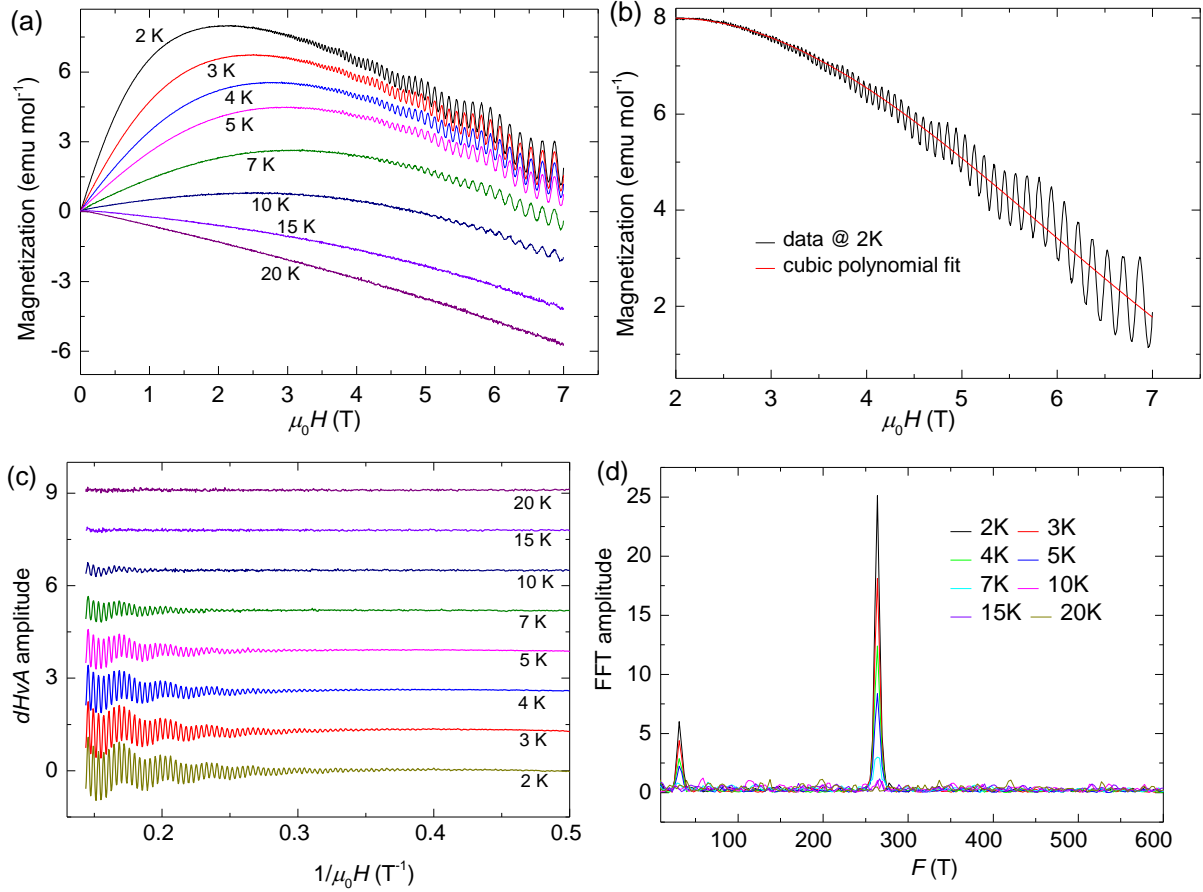

**FIG. S3.** B ∥ [001] and temperature range 2 − 20 K for (a) Isotherm magnetization. (b) Measured magnetization data at 2 K with cubic polynomial fit. (c) AdHv oscillations after background subtracted from (b). (d) Fast furrier transforms of data (c).

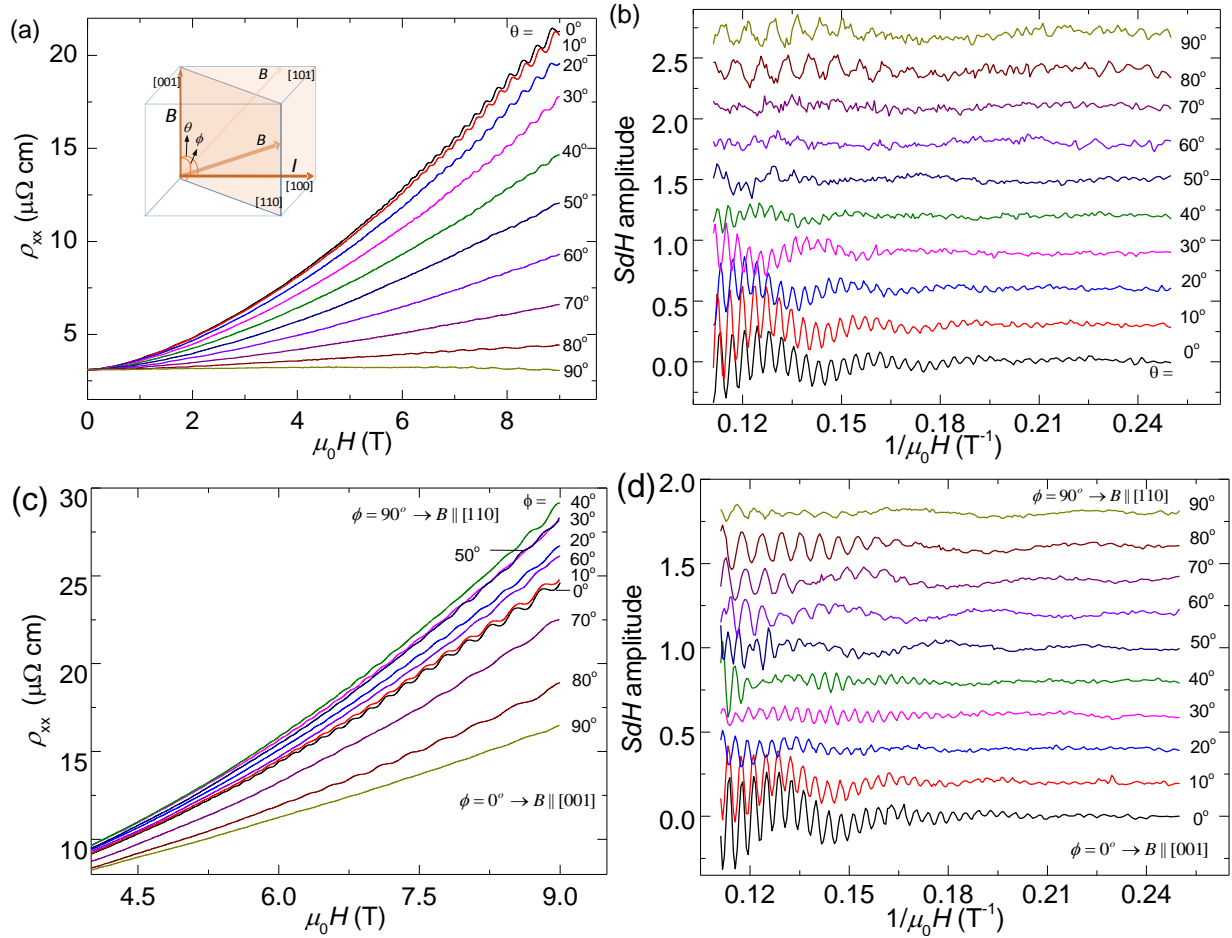

**FIG. S4.** Angle dependent magnetoresistance at 2 K from $\theta, \phi = 0°\ \rightarrow B||[001]$ to (a) $\theta = 90°\ \rightarrow B||[100]$, (b) $\phi = 90°\ \rightarrow B||[110]$. The corresponding FFTs of (a) and (b) after subtraction of a cubic polynomial are in (c) and (d), respectively.